\documentclass[fleqn,10pt]{wlscirep}
\usepackage[utf8]{inputenc}
\usepackage[T1]{fontenc}
\title{QuanTI-FRET: a framework for quantitative FRET measurements in living cells}

\author[1]{Alexis Coullomb}
\author[1]{Cécile M. Bidan}
\author[2]{Chen Qian}
\author[2]{Fabian Wehnekamp}
\author[3]{Christiane Oddou}
\author[3]{Corinne Albigès-Rizo}
\author[2]{Don. C. Lamb}
\author[1,*]{Aurélie Dupont}
\affil[1]{Univ. Grenoble Alpes, CNRS, LIPhy, Grenoble, F-38000, France}
\affil[2]{Department of Chemistry, Center for Nano Science (CENS), Center for Integrated Protein Science (CIPSM) and Nanosystems Initiative München (NIM), Ludwig Maximilians-Universität München, Germany}
\affil[3]{Institute for Advanced Biosciences, Université Grenoble Alpes, INSERM U1209, UMR5309, F38700 La Tronche, France}

\affil[*]{aurelie.dupont@univ-grenoble-alpes.fr}

\begin{abstract} 
 Förster Resonance Energy Transfer (FRET) allows for the visualization of nanometer-scale distances and distance changes. This sensitivity is regularly achieved in single-molecule experiments \textit{in vitro} but is still challenging in biological materials. Despite many efforts, quantitative FRET in living samples is either restricted to specific instruments or limited by the complexity of the required analysis. With the recent development and expanding utilization of FRET-based biosensors, it becomes essential to allow biologists to produce quantitative results that can directly be compared. Here, we present a new calibration and analysis method allowing for quantitative FRET imaging in living cells with a simple fluorescence microscope. Aside from the spectral crosstalk corrections, two additional correction factors were defined from photophysical equations, describing the relative differences in excitation and detection efficiencies. The calibration is achieved in a single step, which renders the Quantitative Three-Image FRET (QuanTI-FRET) method extremely robust. The only requirement is a sample of known stoichiometry donor:acceptor, which is naturally the case for intramolecular FRET constructs. We show that QuanTI-FRET gives absolute FRET values, independent of the instrument or the expression level. Through the calculation of the stoichiometry, we assess the quality of the data thus making QuanTI-FRET usable confidently by non-specialists.

\end{abstract}
\begin{document}

\flushbottom
\maketitle

\thispagestyle{empty}

\section*{Introduction}
The theory behind Förster Resonance Energy Transfer (FRET) was first successfully described in 1946 but its application to biological systems, particularly in living cells, has only become popular in the late 1990s with the cloning of fluorescent proteins. Since the first cloning of the Green Fluorescent Protein (GFP), fluorescence microscopy has rapidly become a standard tool in cell biology. Fluorescence labelling allows the localization of a protein of interest in space and time in a biological specimen, from cells to animals. The labelling of several proteins in the same sample has been used to address protein-protein interactions in terms of colocalization. However, using standard fluorescence microscopy, determination of protein-protein distance is limited by the diffraction of light i.e., hundreds of nanometers. Förster Resonance Energy Transfer (FRET) methods circumvent this barrier by allowing the detection of distances below 10 nanometers between a donor fluorophore and an acceptor through non-radiative energy transfer mediated by dipole-dipole interactions. FRET measurements can distinguish between two proteins being in the same compartment or in direct contact. Moreover, the ability to measure nanometric variations allows for the detection of protein conformational changes \cite{ha_single-molecule_1999,weiss_fluorescence_1999, erickson_preassociation_2001}. A large class of fluorescent biosensors have been engineered based on FRET to monitor protein function (kinase\cite{zhang_genetically_2001,ting_genetically_2001}, GTPase\cite{pertz_spatiotemporal_2006}), calcium signals, \cite{miyawaki_fluorescent_1997} or more recently, forces on the molecular scale \cite{grashoff_measuring_2010, meng_fluorescence_2008, ringer_multiplexing_2017}. The most common design relies on a molecular recognition element coupled with two fluorescent proteins (FPs) expressed in the same amino-acid sequence (intramolecular FRET sensor). An intermolecular FRET design is also possible where the FPs are inserted on two independent moieties. In this case, the apparent stoichiometry can strongly vary, which makes a quantitative analysis much more difficult.

There are two main approaches for measuring FRET in living cells: one is based on the change in fluorescence intensity and the other on the change in the donor fluorescence lifetime \cite{padilla-parra_fret_2012}. Fluorescence LIfetime Microscopy (FLIM) requires sophisticated instrumentation and analysis, and is often recognized as a quantitative method for live-cell measurements. Different strategies have been developed to measure FRET efficiency via the fluorescence intensity of the donor and/or of the acceptor, some involving the total photobleaching of one fluorophore or specific instruments for spectral imaging \cite{chen_characterization_2007,wlodarczyk_analysis_2008,arsenovic_sensorfret:_2017}. The most compatible method with dynamic quantitative FRET imaging and live-cell imaging is based on the sensitized-acceptor emission. Because the collected fluorescence intensity depends strongly on numerous instrumental factors (excitation, filter set, camera sensitivity etc), this approach requires several corrections to calculate an instrument-independent FRET efficiency. The literature is rich of different correction factors and mathematical expressions of  FRET indices \cite{ berney_fret_2003,zeug_quantitative_2012}. The idea of correcting for spectral crosstalks and at least for the difference in detection efficiency between donor and acceptor channels emerged concomitantly in the single-molecule \cite{dahan_ratiometric_1999, kapanidis_fluorescence-aided_2004}  and in the live-cell imaging fields \cite{youvan_fluorescence_1997, gordon_quantitative_1998, zal_photobleaching-corrected_2004}. It is now generally accepted that bleedthrough of the donor emission in the acceptor channel and direct excitation of the acceptor by donor excitation channel must be corrected by substracting their contributions. This requires the acquisition of three different signals, also called 3-cube strategy in live-cell imaging \cite{gordon_quantitative_1998}. As such, the apparent FRET index varies with the fluorophore concentration and, even with additional normalization, the direct comparison of FRET values obtained independently is not possible\cite{hochreiter_advanced_2019}. To account for photophysical artifacts, we need to go back to physical equations and determine the origin of the signal in each channel. The next obstacle is the experimental determination of the correction factors. Existing methods require samples with known FRET efficiency \cite{hoppe_fluorescence_2002} or known concentration \cite{thaler_quantitative_2005} or even an additional experiment using acceptor photobleaching \cite{zal_photobleaching-corrected_2004}.
 
In this work, we clarify the theory coming from single-molecule studies\cite{lee_accurate_2005} and adapt it to live-cell imaging. We present a new method to determine all the correction factors in a robust manner without any additional photobleaching experiment or external calibration of the FRET efficiency. The only requirement for calibration is knowledge of the donor:acceptor stoichiometry, which is in general known by construction. The calibration can thus be achieved directly on the sample of interest or with FRET standards\cite{koushik_cerulean_2006}. While the stoichiometry can be accurately measured in the last case, this information can always be used as a quality factor to discard aberrant pixels. No specialized microscope is required as the QuanTi-FRET (Quantitative
Three-Image) method can be applied to any epifluorescence triplet of images acquired with commercial instruments. Here, we demonstrate that QuanTI-FRET allows for absolute FRET measurements that are independent of the instrumental setup and of the fluorophore concentration. Being robust and including an inherent data quality check, the method can be used confidently by non-specialists, especially for FRET-based biosensors applications.

\section*{Theory}

\begin{figure}[tbp]
\centering
\includegraphics[width=16cm]{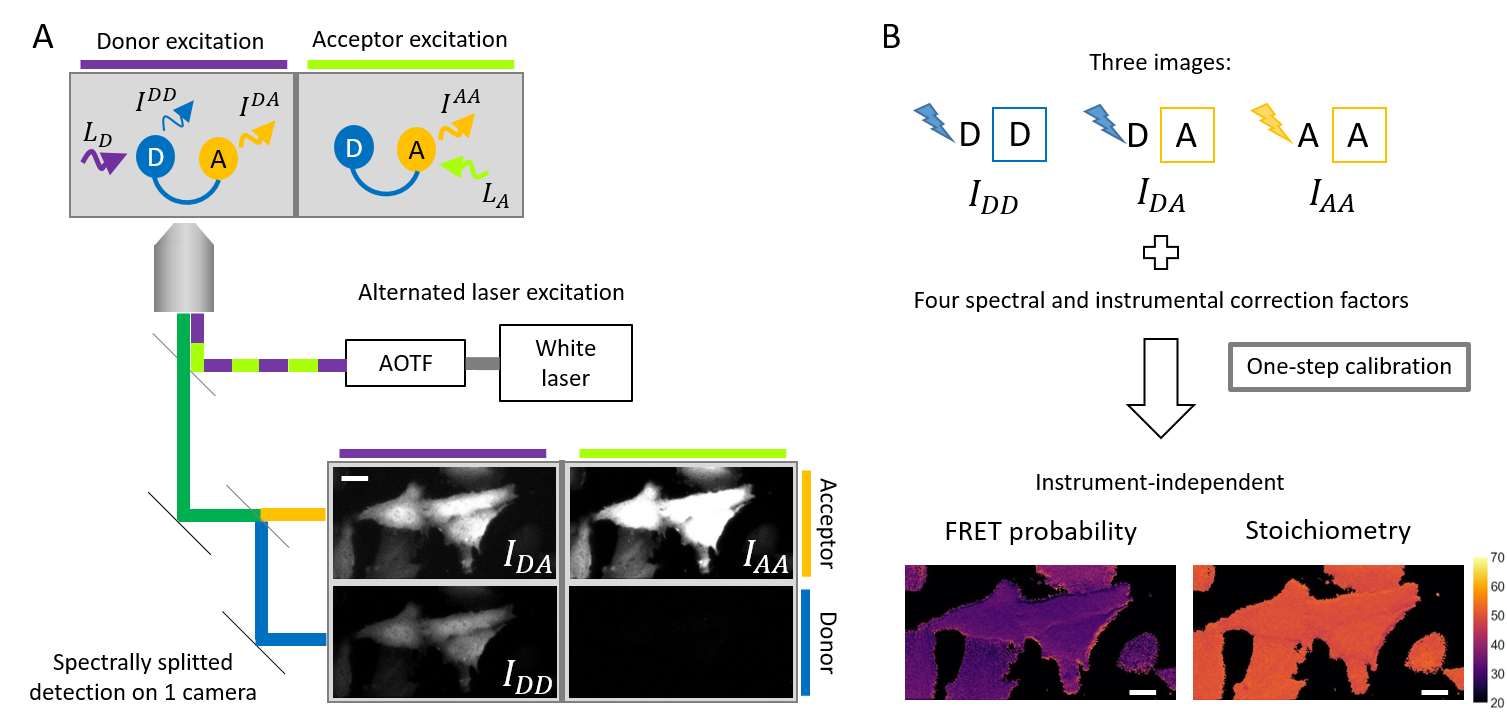}
\caption{ A. A schematic of an experimental setup used for the validation of the framework is shown. Three images are acquired in two snapshots by automatically alternating the laser excitation and splitting the camera in two detection channels corresponding the donor and acceptor channels. B. Framework for quantitative FRET analysis. The analysis requires three images combining the detection in the donor and the acceptor channels with the excitation of the donor and the acceptor. A calibration step allows the determination of four factors correcting for the crosstalks and the relative excitation and detection efficiencies of the donor and acceptor fluorophores. As a result, instrument-independent FRET probabilities and stoichiometries are calculated. Scale bar: 20$\mu$m.}\label{Fig1}
\end{figure}

To obtain as much information as possible from the sample, we follow a multiple excitation scheme (Fig. \ref{Fig1}) as introduced by Kapanidis and colleagues for single molecule spectroscopy \cite{kapanidis_fluorescence-aided_2004} and close to the three-cube method in live-cell imaging\cite{youvan_fluorescence_1997}. By switching rapidly between both excitation sources, and splitting the emission into two channels on the camera, we acquire in two successive snapshots four images:

\begin{description}
		\item [$I_{DD}$]: the detected signal in the donor channel after excitation at the donor wavelength,
		\item [$I_{DA}$]: the detected signal in the acceptor channel after excitation at the donor wavelength,
		\item [$I_{AA}$]: the detected signal in the acceptor channel after excitation at the acceptor wavelength.
\end{description}

The fourth image $I_{AD}$ contains no information, only noise, and is discarded. In principle, only $I_{DD}$ and $I_{DA}$ are sufficient to calculate the transfer efficiency. That would be the case if the photons coming from the donor and the acceptor had the same detection efficiency. In practice, it is not possible to have such an instrument and, several corrections must be considered to get unbiased quantitative FRET efficiencies. The third image, $I_{AA}$, is independent from the FRET efficiency but is required to calculate all the necessary corrections.

One can write the intensity of the three types of signals as a function of the photophysical and instrumental parameters and the population of the different fluorophores, $n^A$ (acceptor) and $n^D$ (donor), and FRET probability, $E$: 

	\begin{align} 
I_{AA} &= n^A L_{A} \sigma_{Aex}^A \phi_A \eta_{Adet}^{Aem} \label{IAA}\\
I_{DD} &= n^D L_{D} \sigma_{Dex}^D (1-E) \phi_D \eta_{Ddet}^{Dem} \label{IDD}\\
I_{DA} &= n^D L_{D} \sigma_{Dex}^D E \phi_A \eta_{Adet}^{Aem} + n^D L_{D} \sigma_{Dex}^D (1-E) \phi_D \eta_{Adet}^{Dem} + n^A L_{D} \sigma_{Dex}^A \phi_A \eta_{Adet}^{Aem} \label{IDA}
 	\end{align}

\noindent where $L_{i}$ is the excitation intensity at the wavelength chosen for excitation of fluorophore $i$, $\sigma_i^j$ is the absorption cross section of $j$ at the excitation wavelength of $i$, $\phi_i$ is the quantum yield of $i$ and $\eta_i^j$ is the detection efficiency of photons emitted by $j$ in the detection channel $i$. The expression of $I_{AA}$ is the simplest as it only depends on species $A$ (acceptor). For $I_{DD}$, one has to take into account the probability to transfer energy to the acceptor, $E$, as acceptor photons are not detected in this channel. Finally, to express $I_{DA}$, the FRET image, not only the signal coming from FRET events must be taken into account but also the two crosstalk terms: (i) the bleedthrough of photons emitted by the donor into the acceptor channel and (ii) the direct excitation of acceptor molecules with the donor specific wavelength. Some of the parameters in the above equations are difficult to measure. We follow a pragmatical approach and avoid the systematic determination of all twelve unknowns. First, we can simplify the expressions by defining a bleedthrough correction factor as $\alpha^{BT}$ and a direct excitation correction factor as $\delta^{DE}$. Additionally, a correction factor for the different detection efficiencies in both channels is defined as $\gamma^M$, and similarly  a correction factor for the different excitation efficiencies in both channels is defined as $\beta^X$
 
\begin{equation}
    \alpha^{BT} = \dfrac{\eta_{Adet}^{Dem}}{\eta_{Ddet}^{Dem}}  \quad \delta^{DE} = \dfrac{L_{D}\sigma_{Dex}^A}{L_{A}\sigma_{Aex}^A} \quad   \gamma^M=\dfrac{\phi_A \eta_{Adet}^{Aem}}{\phi_D \eta_{Ddet}^{Dem}} \quad
    \text{and} \quad \beta^X=\dfrac{L_{A} \sigma_{Aex}^A}{L_{D} \sigma_{Dex}^D}
\label{corrfact}
\end{equation} 
 
 Hence, the notation is simplified and by inverting the previous set of equations, we obtain the FRET probability:
\begin{equation}
	E = \dfrac{I_{DA}-\alpha^{BT}I_{DD}-\delta^{DE} I_{AA}}{I_{DA}-\alpha^{BT}I_{DD}-\delta^{DE} I_{AA}+\gamma^M I_{DD}} \label{E1}    
\end{equation}

In addition, as in Lee et al. \cite{lee_accurate_2005}, we define the stoichiometry as the relative amount of donor molecules with respect to the total number of fluorophores in each pixel:
\begin{equation}
    S = \dfrac{n^D}{n^D+n^A} 
\label{S1}
\end{equation}
From equations (\ref{IAA}) and (\ref{IDD}), we derive expressions for $n^D$ and $n^A$ and insert them into equation (\ref{S1}). By simplifying with the excitation correction factor $\beta^X$ defined in equations (\ref{corrfact}), equation (\ref{S1}) reduces to:

\begin{equation}
S = \dfrac{1}{1+\dfrac{I_{AA}}{I_{DD}} \dfrac{1}{\beta^X\gamma^M}(1-E)}
\end{equation}
To decouple stoichiometry and FRET probability, we replace $E$ by the expression given in equation (\ref{E1}). Finally the stoichiometry reads:
\begin{equation}\label{S3}
S = \dfrac{I_{DA}-\alpha^{BT}I_{DD}-\delta^{DE} I_{AA}+\gamma^M I_{DD}}{I_{DA}-\alpha^{BT}I_{DD}-\delta^{DE} I_{AA}+\gamma^M I_{DD}+I_{AA}/\beta^X}
\end{equation}

By including the crosstalk corrections into a corrected FRET image, $I_{DA}^{corr} = I_{DA}-\alpha^{BT}I_{DD}-\delta^{DE} I_{AA}$, we obtain two master equations defining the FRET probability and the stoichiometry in each pixel:
\begin{align}
	E &= \dfrac{I_{DA}^{corr}}{I_{DA}^{corr}+\gamma^M I_{DD}} \label{Efinal} \\ 
    S &= \dfrac{I_{DA}^{corr} +\gamma^M I_{DD}}{I_{DA}^{corr} +\gamma^M I_{DD}+I_{AA}/\beta^X} \label{Sfinal}
\end{align}
Both E and S can be calculated from the three experimental images, $I_{DD},I_{DA}$ and $I_{AA}$, and four parameters, $\alpha^{BT}, \delta^{DE}, \gamma^{M}$ and $\beta^X$. All four correction factors are derived from the detailed expressions of the collected fluorescence intensities in the three different channels. The notations were chosen according to the consensus in the single-molecule field \cite{hellenkamp_precision_2018} with a supplemental exponent for a direct understanding of the role of each correction factor.
The crosstalk correction factors are already widely used in the 3-cube approaches \cite{zal_photobleaching-corrected_2004} and are straightforward to calibrate. Imaging a donor-only sample, in vitro or in cellulo, provides $\alpha^{BT}$; similarly, imaging an acceptor-only sample provides $\delta^{DE}$. $\alpha^{BT}$ depends only on the donor emission spectrum, the filter set and the spectral response of the camera. $\delta^{DE}$ depends on the acceptor excitation spectrum but also on the ratio of the illumination power in the two channels. Under the same experimental conditions (same fluorophores, same filter set and illumination intensities), the crosstalk corrections to be brought to $I_{DA}$ depend only on the quantity of both fluorophores, given by $I_{DD}$ and $I_{AA}$ while $\alpha^{BT}$ and $\delta^{DE}$ are unchanged.

The two other correction factors,  $\gamma^M$ ("M" for Emission) and $\beta^X$ ("X" for Excitation), are more difficult to determine. $\gamma^M$ accounts for the difference in the measured fluorescence emission when the same number of donor or acceptor molecules are excited. Hence, it is related to the quantum yield and to the detection efficiency of the setup in each channel. $\beta^X$ accounts for the difference in energy absorption for each channel. Hence, it is related to the illumination intensity and the absorption cross-section of each fluorophore. $\gamma^M$  has already been described, in single molecule \cite{dahan_ratiometric_1999} and in live-cell imaging \cite{zal_photobleaching-corrected_2004}. Several indirect strategies have been developed to determine the value of $\gamma^M$: from acceptor photobleaching \cite{ha_probing_1996, zal_photobleaching-corrected_2004 }, the use of a FRET sample with known FRET efficiency\cite{hoppe_fluorescence_2002}, an interpolation from two constructions with very different FRET values \cite{chen_measurement_2006} or a fit of the relation between $1/S$ and $E$\cite{lee_accurate_2005}. $\beta^X$ has been introduced by Lee et al. \cite{lee_accurate_2005} for single molecule experiments and a similar parameter has also been empirically introduced by Chen et al. for cells experiments \cite{chen_measurement_2006}. If $\beta^X$ and $\gamma^M$ are determined independently, $\beta^X$ has no effect on the FRET efficiency but just on the stoichiometry (see equations (\ref{Efinal}) and (\ref{Sfinal})) . Since the stoichiometry in single molecule studies is often limited to donor only, acceptor only and donor:acceptor complexes, S does not need to be accurate and $\beta^X$ is not necessary. On the contrary, we will show that $S$ can be very useful in live-cell experiments even when the FRET construction has a well-defined stoichiometry.

\subsection*{Calibration of the correction factors}
Having described the theory directly from the physical parameters of the fluorophores and of the experimental setup, the difficult part to achieve the calculation of quantitative FRET is to determine the four correction factors. As mentioned previously, the crosstalk correction factors are measured from donor-only and acceptor-only cells, and calculated as the ratios 
\begin{equation}
    \alpha^{BT}=\dfrac{I_{DA}^{donor-only}}{I_{DD}^{donor-only}}  \quad \text{and} \quad \delta^{DE}=\dfrac{I_{DA}^{acceptor-only}}{I_{AA}^{acceptor-only}} \label{BTDE}
\end{equation}
These ratios are calculated in each pixel of all the imaged cells and the median value is kept. 
The correction factors $\gamma^M$ and $\beta^X$ cannot be determined from the donor-only or acceptor-only samples where the FRET probability is equal to zero (or not defined). Another piece of information is necessary and is found in the stoichiometry. Equation (\ref{Sfinal}) can be rewritten as 
\begin{equation}
    \beta^X \gamma^M I_{DD} + \beta^X I_{DA}^{corr} = \dfrac{S}{1-S} I_{AA}, \label{eqplane}
\end{equation}

which is the equation of a plane in the 3D space defined by \{$I_{DD},I_{DA}^{corr},I_{AA}$\}. If the stoichiometry is known, the strategy is to fit the experimental data \{$I_{DD},I_{DA}^{corr},I_{AA}$\} to a plane and thereby determine $\beta^X \gamma^M$ and $\beta^X$. If the FRET sample of interest has an unknown stoichiometry, another calibration experiment has to be made with a defined stoichiometry FRET probe.  
Practically, the pixel values of a whole dataset (N cells) are gathered in vectors $X=[I_{DD},I_{DA}^{corr}]$ and $Y=[I_{AA}]$ and the matrix $A=[\gamma^M \beta^X, \beta^X]$ is determined such as $X.A = Y$ by a least-square fitting. If the sample shows only one FRET value $E$ with different fluorescence intensities (i.e. fluorophore concentrations), the pixel values will form a straight line in the 3D space \{$I_{DD},I_{DA}^{corr},I_{AA}$\} (Fig.\ref{Fig2}). As a result, an infinite number of planes can fit the dataset. For a good determination of  $\beta^X$ and $\gamma^M$, it is therefore necessary that the FRET values of the dataset are sufficiently spread. The visualization and the calculation of the correction factors in the 3D space \{$I_{DD},I_{DA}^{corr},I_{AA}$\} is the originality of this work. We compare our approach with the two other related methods in the last section.

\section*{Results}
\subsection*{Validation of QuanTI-FRET using FRET standards}

\begin{figure}[tbp]
\centering
\includegraphics[width=15cm]{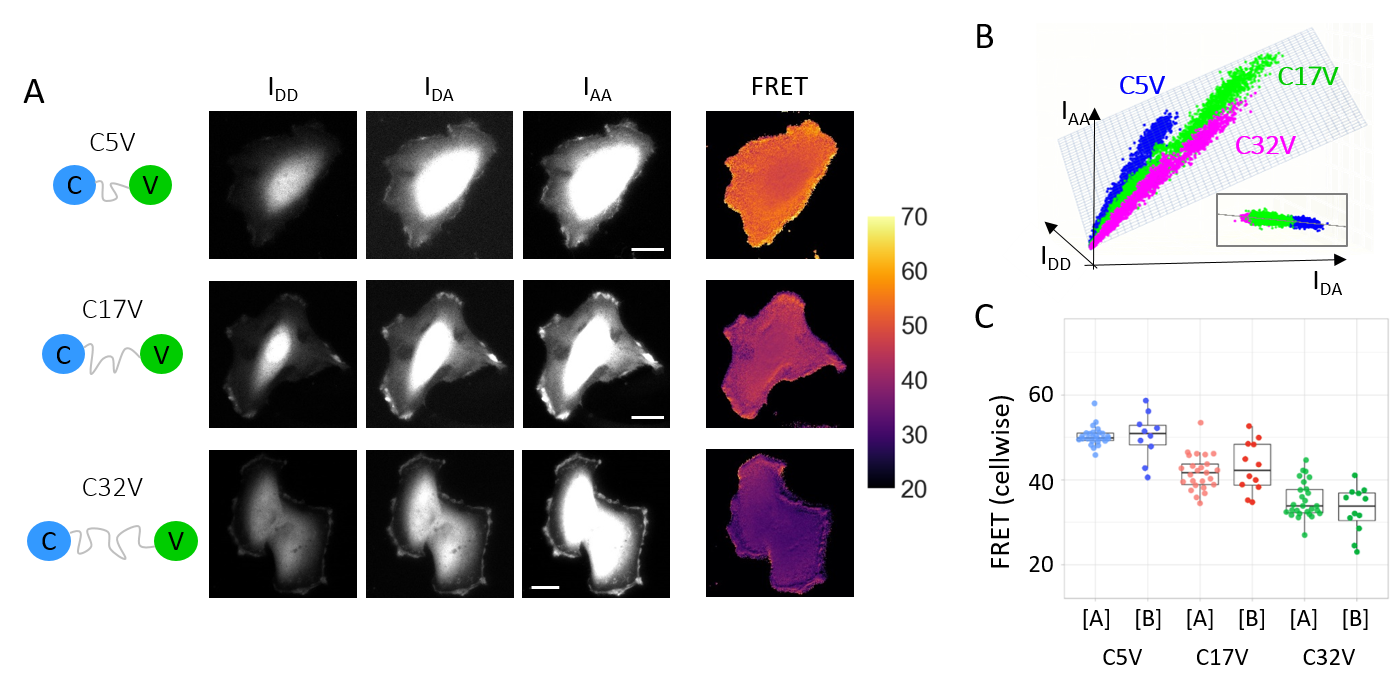}
\caption{FRET measurements on the three FRET standards, C5V, C17V and C32V. (\textbf{A}) Triplet fluorescence images are shown for exemplary cells transfected with the three FRET standards: C5V (short linker), C17V (medium linker) and C32V (long linker). The calculated FRET maps for the individual cells are shown on the right plotted using the same color scale. The highest FRET is observed for the shortest linker construct C5V and decreases to the lowest FRET construct C32V. Scale bar: 20$\mu$m. Color bar: FRET efficiency in \% (\textbf{B}) Scattered plot of all pixels values from all cells imaged in the \{$I_{DD},I_{DA}^{corr},I_{AA}$\} 3D-space and the fitted plane, side view as inset. The three FRET standard populations forming three distinct clouds are all lying on the plane defined by $\beta^X$ and $\gamma^M$. (\textbf{C}) Boxplot gathering cellwise FRET values of C5V, C17V and C32V measured independently in two different labs ([A] and [B]). After calibration, the same FRET median values were obtained.}\label{Fig2}
\end{figure} 

To test the proposed method in live-cell experiments, we utilized the FRET standards developed by Thaler et al. \cite{thaler_quantitative_2005} and Koushik et al. \cite{koushik_cerulean_2006}. The FRET standards consist of a pair of fluoroscent proteins, a donor (Cerulean) and an acceptor (Venus), separated by an amino-acid sequence of variable length. Three standards were used in the present work to calibrate the experimental setup:  C5V, C17V and C32V, where the linker between donor and acceptor consisted of 5, 17 and 32 amino-acids respectively. The construct with the shortest linker, C5V, was expected to exhibit the largest FRET efficiency and the FRET efficiency to decrease as the linker length increases \cite{koushik_cerulean_2006}. The FRET standards were expressed in Hela cells and imaged on the setup described in Figure\ref{Fig1}.

As a first step for calibration, the crosstalk corrections corresponding to the donor, Cerulean, and the acceptor, Venus, must be determined. Hence, Cerulean-only cells and Venus-only cells were imaged. Using equation (\ref{BTDE}), the bleedthrough for Cerulean was calculated as $\alpha^{BT} = 0.421 \pm 0.002$ (10 cells) and the direct excitation of Venus as $\delta^{DE} = 0.1100 \pm 0.0008$ (12 cells). The  pixelwise distributions of $\alpha^{BT}$ and $\delta^{DE}$ are shown in supplementary information (Fig.S1).
The second step consists in the determination of $\gamma^M$ and $\beta^X$, the factors correcting for the difference in detection and excitation efficiencies in the different channels. The three necessary fluorescence images, $I_{DD}, I_{DA}$ and $I_{AA}$, of three exemplary cells transfected with C5V, C17V and C32V  are shown in Figure \ref{Fig2}A. All the pixel values \{$I_{DD},I_{DA},I_{AA}$\} of all cells expressing the three constructs were gathered as one dataset and fitted with the plane equation (\ref{eqplane}) (Fig.\ref{Fig2}B). A mask of each cell was obtained and only the pixels coming from within the cells were kept. This equation has an additional unknown, S. An assumption on S is necessary at this step.  By design, the CxV constructs should have on average  one donor for one acceptor, \textit{i.e.} $S=0.5$. This assumes the maturation efficiency of the donor and the acceptor are close to 1. We will discuss the influence of maturation in the next section. For S=0.5, the plane equation reduces to: 
\begin{equation}
    \beta^X \gamma^M I_{DD} + \beta^X I_{DA}^{corr} =  I_{AA}. \label{eqplane05}
\end{equation}
A given set of experimental conditions (laser power, filter set, fluorophores, stoichiometry) corresponds to one plane, and in this plane, a given FRET efficiency corresponds to a line. As seen in Fig.\ref{Fig2}B, the scattered data from the three standards appear as linear clouds lying on the same plane defined by $\beta^X$ and $\gamma^M$ and the assumed stoichiometry $S$ ($S=0.5$, 1 donor:1 acceptor). Here, the least square fitting of the plane yielded $\beta^X= 1.167 \pm 0.008$ and $\gamma^M = 2.10 \pm 0.02$ with a coefficient of determination $R^2 = 0.995$.

Once all the correction factors are determined, the FRET probability can be measured. Since this dataset was used for calibration with the hypothesis of $S=0.5$, the stoichiometry cannot be an output for this calibration dataset. Nevertheless, no assumption was made concerning $E$, and therefore, the FRET probability can be calculated on the same dataset as the one used for calibration. If the experiment of interest presents a sufficiently broad distribution of FRET probabilities to determine the plane in 3D, there is no need for a different experiment with FRET standards for calibration. Hence, calibration can be achieved on-the-fly on samples with known stoichiometry.

More than 25 Hela cells expressing one CxV construct were measured. The median FRET probability was $E_{C5V} = 51.1$ ($s.d. = 12.2$, 26 cells) for C5V, $E_{C17V} = 43.1$ ($s.d. = 11.8$, 25 cells) for C17V and $E_{C32V} = 35.1$ ($s.d. = 11.5$, 27 cells) for C32V, calculated over more than $3 \cdot 10^{6}$ pixels. The uncertainty comes rather from the cell to cell variability than from the pixel statistics. Hence, the median FRET value per cell was taken (Fig.\ref{Fig2}C, dataset [A]) and the uncertainty calculated as the standard error of the mean yielding: $E_{C5V} = 50.3 \pm 0.4$, $E_{C17V} = 41.7 \pm 0.8$ and $E_{C32V} = 35.1 \pm 0.8$. To verify that the FRET probability was independent of the fluorescence intensity, the Spearman's rank correlation coefficient was calculated between $E$ and $I_{AA}$, the only channel not affected by FRET and just related to the fluorophore concentration. Gathering the data from all three standards, the resulting Spearman's coefficient was $\rho = 0.04$, confirming the absence of correlation between the fluorescence level and the calculated FRET probability. This is also true pixelwise on a single cell basis (see Supplementary Fig.S2) and cellwise comparing all cells expressing one FRET standard (see Supplementary Fig.S3). Similarly, we questioned the effect of the correction factor $\gamma^M$ by calculating the Spearman's coefficient between $E$ and the total donor fluorescence ($\gamma^M I_{DD} + I_{DA}^{corr}$) without the correction, $\rho = 0.111$ ($\gamma^M=1$), and with the correction $\rho = 0.045$ ($\gamma^M=2.10$). Hence, correcting for the different detection efficiencies decreases the correlation by a factor 2.5 between the donor fluorescence and the calculated FRET probability.

The goal of the QuanTI-FRET method is to enable the comparison of FRET-based experiments from different studies \textit{i.e.}, obtained independently in different laboratories in the world. To test this, we performed the same experiments a second time in a completely independent way: with a different instrument, in a different country, by a different team on another cell culture with fresh constructs ordered directly from Addgene. The experimental data was analyzed with the exact same procedure. The calibration gave the following correction factors  $\alpha^{BT} = 0.467\pm 0.001$ (12 cells) and $\delta^{DE}=0.101 \pm 0.003$ (12 cells)  for the crosstalks and  $\beta^X = 2.03 \pm 0.07$ and $\gamma^M = 1.35 \pm 0.07$ ($R^2=0.82$) for the excitation and emission  corrections. The FRET probability was measured for the three FRET standards giving $E_{C5V} = 50 \pm 1.7$ (10 cells), $E_{C17V} = 43 \pm 1.6$ (12 cells) and $E_{C32V} = 33 \pm 1.5$ (12 cells) (Fig.\ref{Fig2}C, dataset [B]). For an easier comparison, correction factors and FRET probabilities from lab [A] and [B] are gathered in Supplementary Materials (Table S1). The variability in this second dataset was larger as seen by the smaller coefficient of determination ($R^2 = 0.82$) of the 3D fitting and the standard deviation of the FRET probability for each construct. Nevertheless, the FRET values obtained were in agreement with the first dataset ([A]). Hence, we show that measuring FRET with the QuanTI-FRET method is quantitative: the absolute FRET values are meaningful and can be compared from one lab to another.

\subsection*{Taking advantage of S}
\begin{figure}[tbp]
\centering
\includegraphics[width=15cm]{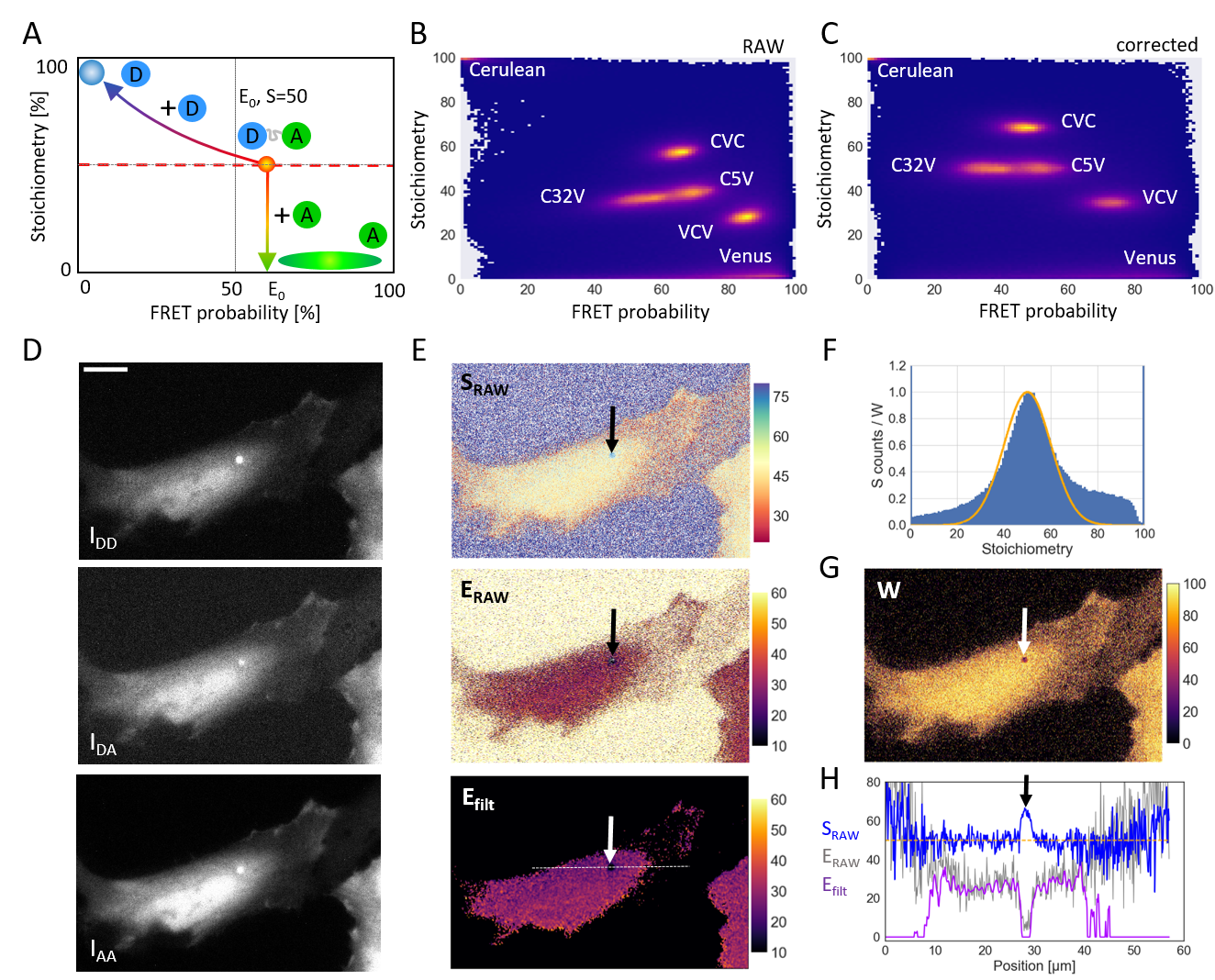}
\caption{(\textbf{A}) Influence of free donor or free acceptor in the sample. Theoretical S-E histogram with trajectories corresponding to the addition of free donor or free acceptor to a construct with 1:1 donor to acceptor ratio. (\textbf{B}) Experimental histogram of S versus E for constructs showing different FRET values (C32V and C5V) or different stoichiometries (CVC and VCV) as well as pure donor (Cerulean) and pure acceptor (Venus). This histogram was calculated using only the crosstalk correction. (\textbf{C}). The same experimental E-S histogram with the complete calibration including $\gamma^M$ and $\beta^X$. In the completely corrected 2D histogram, the stoichiometry and FRET probability are uncorrelated ($\rho = 0.02$). (\textbf{D}) Exemplary triplet of images showing a cell expressing C32V with a low signal-to-noise ratio, Scale bar 15$\mu$m. (\textbf{E}) The corresponding RAW E and S maps and the FRET map for the images in panel \textbf{D} after filtering with the weigthed gaussian filter. (\textbf{F}) The corresponding stoichiometry histogram and the weights (W) as a function of the stoichiometry (line). The weights are given, for each pixel, by a gaussian function of the deviation from the expected stoichiometry ($S=0.5$) with a variance $\sigma_S = 0.1$. The corresponding map of weights W is shown in (\textbf{G}). (\textbf{H}) Line profiles corresponding to the three maps shown in panel \textbf{E}. Due to high intensity background in an endosome, the FRET efficiency drops (thin grey line). This anomaly is also observable in the stoichiometry (blue). By weighting the image with the measured stoichiometry, such artifacts can be avoided (magenta).}.\label{Fig3}
\end{figure}

So far, the stoichiometry was used only to calibrate the system. However, once the experimental system has been calibrated, the QuanTI-FRET analysis can determine both E and S independently. In this case, additional information can be extracted from S. First of all, S can be used to evaluate the quality of the calibration and of the dataset. As in single molecule studies, the 2D histogram combining the stoichiometry and FRET probability histograms (Fig.\ref{Fig3}A) is a useful tool . In theory, the standard constructs with 1 donor for 1 acceptor should appear as a cloud corresponding to their average FRET efficiency, $E_0$, and $S=0.5$. A known stoichiometry of 1:1 donor:acceptor is also reasonable for a biosensor construct that contains both donor and acceptor fluorescent proteins that fold and mature with high efficiency. However, when looking for interactions between different proteins, a fraction of donor only and/or acceptor only constructs are expected. If free acceptors are also present in the image, the apparent FRET probability stays constant but the stoichiometry drops (Fig.\ref{Fig3}A). On the contrary, if free donor is present with the 1:1 construct, both S and E are affected. This variation can be described theoretically. If a solution containing a donor-acceptor construct, $n_D^0$, with an average FRET efficiency of $E_0$ is mixed with free donor, $n_D^{free}$, the apparent FRET probability and the apparent stoichiometry are (see Supplementary Information):
\begin{equation}
E_{app} = \dfrac{1}{1+\dfrac{1-E_0 + n^D_{free} /n^D_0}{E_0}} \quad \text{and} \quad 
S_{app} = \dfrac{1+n^D_{free} /n^D_0}{1/S_0 + n^D_{free} /n^D_0} \label{Eap}
\end{equation}
We can then write the analytical formula describing this mix in the E-S histogram:
\begin{equation}
    S_{app} = \dfrac{E_0 / E_{app}}{1/S_0 + E_0 / E_{app} - 1}, \label{ESap}
\end{equation}
which is sketched in Fig.\ref{Fig3}A. In equations (\ref{IDD}) and (\ref{IDA}), we assumed that all the donors were able to FRET \textit{i.e.}, had an acceptor partner. If this is not the case and free donors exist, then E becomes an apparent FRET probability $E_{app}$ as in equation (\ref{Eap}). If the experimental E-S histogram can be fitted to equation (\ref{ESap}), the FRET probability, $E_0$ of the 1:1 construct can be extracted. The presence of free donors can result from the poor efficiency of the acceptor fluorophore to fold. As demonstrated above, this case can easily be seen and treated with the QuanTI-FRET method. The presence of free acceptors does not affect the FRET efficiency once the system calibrated. If free acceptors are present in the calibration samples, one should at least evaluate and take into account the effective stoichiometry in order to obtain a reliable calibration and avoid the propagation of  biases to the measurements of interest. If both free donors and free acceptors are present, the situation is more complicated due the ensemble measurement made in each pixel. But fortunately, most of FRET-based biosensors are formed with variants of GFP, in particular of the pair CFP/YFP, which fold well \cite{frommer_genetically_2009,sizaire_fret-based_2017}.

The observation of the E-S 2D histogram gives a hint about the quality of the calibration. In theory, for a sample with a fixed stoichiometry, the FRET probability and the stoichiometry should be uncorrelated resulting in horizontal clouds in the 2D histogram. Figure \ref{Fig3}B shows the experimental data from this work with crosstalk correction but with $\beta^X$ and $\gamma^M$ both set equal to 1. The constructs C5V and C32V do not lie on a horizontal line whereas they should have the same stoichiometry. On the contrary, with the complete calibration of  $\beta^X$ and $\gamma^M$ (Fig. \ref{Fig3}C), the two clouds lie on a horizontal line corresponding to $S=0.5$ (Spearman's correlation coefficient between E and S: $\rho=0.02$ for C5V-C17V-C32V).

Once the system has been calibrated with FRET probes with a known stoichiometry, the stoichiometry becomes an output of the QuanTi-FRET analysis. Two additional FRET standards were imaged under the same conditions as before, CVC (2 donors:1 acceptor) and VCV (1 donor:2 acceptors) \cite{thaler_quantitative_2005}. As these two constructs were not used to determinate $\gamma^M$ and $\beta^X$, no assumption was made with respect to their stoichiometry. Both constructs were built with the same fluorophore pair and imaged using the same conditions (filter set, laser power, camera), hence, the calibration was still valid. Practically, the experimental results gave $S=68.9 \pm 0.2$ for CVC (24 cells), with an expected value of $66\%$, and $S=35.1 \pm 0.2$ for VCV (9 cells), with an expected value of $33\%$ (Supplementary Fig.S4). Importantly, CVC and VCV experiments are well calibrated, appearing as horizontal clouds in the E-S histogram (Fig.\ref{Fig3}C). On the same histogram, the acceptor (Venus) population was found at very low stoichiometry ($S= 12 \pm 4$, 12 cells) as expected and the donor population is also found where expected at stoichiometry close to 1 ($S= 98.7 \pm 0.4$, 10 cells). 

In the case of a fixed stoichiometry sample, as is the case for most FRET-based biosensors, S can still bring an important piece of information about the confidence. The usual way to determine the uncertainty about a pixel is to rely on the photon statistics: if the fluorescence signal is high, then a high confidence is assumed. This is certainly true for pure fluorescence imaging but, in the case of FRET, there are cases where a high fluorescence intensity occurs in pixels where the FRET is biased. For instance, FRET can be affected by the local chemical environment (pH), the local crowding or by any unequal effect on the fluorescence of the donor and acceptor. An example is shown on Figure \ref{Fig3}D where lower-than-expected FRET efficiency was observed in certain bright intracellular vesicles. The corresponding raw results of the pixel-based analysis is shown in Fig.\ref{Fig3}E ($S_{RAW}$ and $E_{RAW}$) and line profiles are plotted (Fig.\ref{Fig3}H). For this example, the spot pointed to by the arrow has a high fluorescence intensity in the three channels but the stoichiometry differs from the expected 50\% (close to 65\%). Similarly, dark, out-of-cell regions of the image also show deviation from the expected stoichiometry. We define a confidence index $W$ as:
\begin{equation}
    W = e^{-\dfrac{(S-S_0)^2}{2 \sigma_S ^2}},\label{eqW}
\end{equation}

where $S_0$ is the expected S and $\sigma_S$ is a parameter to tune the sensitivity. $W$ renders the deviation from an expected stoichiometry as a score between 0 and 1 ($S=S_0$) with a gaussian shape (Fig.\ref{Fig3}F). This confidence index can be used directly to display FRET maps with color-coded FRET values and brigthness-coded $W$. To go one step further, the confidence index can be inserted in a spatial filter. Indeed, FRET maps often need to be spatially averaged, the actual resolution being limited by the diffusion of the FRET species and larger than the pixel size. A weighted gaussian filter was therefore designed where the effect of a gaussian kernel (typically 7x7 pixels$^2$) was locally weigthed with $W$ (Fig.\ref{Fig3}G) as follows:
\begin{equation}
    E_{filt} = \frac{(W \circ E)*G}{W*G},
\end{equation}
where $*$ denotes a convolution and $\circ$ the Hadamard product, $E$ and $W$ are dealt as matrices corresponding to the raw FRET image and the weights as defined in equation (\ref{eqW}), $E_{filt}$ being the filtered FRET map. As the gaussian distribution never reaches zero, an additional threshold was applied based on the local weight of the considered pixel. An example is shown on Figure\ref{Fig3}E, the application of the weighted gaussian filter ($\sigma_S=0.1 , \sigma_{Gauss}=1.5$ and threshold on W $W_{th}=0.5$) totally eliminates the background around the cells and also very dim areas inside cells as well as the bright vesicle with anomalous stoichiometry (Fig.\ref{Fig3}H).

\section*{Discussion}
\begin{table}[tbp]
    \centering
    \begin{tabular}{|l|l|l|l|l|l|}
    \hline
           & $\beta^X$ & $\gamma^M$ & C5V & C17V & C32V \\
          \hline
          QuanTI-FRET & $1.167 \pm 0.008$ & $2.10 \pm 0.02$ & $50.3 \pm 0.4$ & $41.7 \pm 0.8$ & $35.1 \pm 0.8$ \\
          \hline
          Lee et al.\cite{lee_accurate_2005} & $1.13 \pm 0.01$ & $2.37 \pm 0.05$ & $47.5 \pm 0.4$ & $38.9 \pm 0.8$ & $32.4 \pm 0.8$\\
          \hline
          Chen at al.\cite{chen_measurement_2006} & $1.135 \pm 0.005$ & $2.19 \pm 0.02$ & $49.4 \pm 0.4$ & $ 40.8 \pm 0.8$ & $34.2 \pm 0.8$\\
          \hline
    \end{tabular}
    \caption{Systematic comparison of QuanTI-FRET method with previous work from Lee et al.\cite{lee_accurate_2005} and Chen et al.\cite{chen_measurement_2006} Dataset [A] was analyzed with the three methods, the resulting correction factors and FRET probabilities for C5V , C17V and C32V are given in this table, with the uncertainty on $\beta^X$ and $\gamma^M$ resulting from a different bootstrap analysis.}
    \label{table1}
\end{table}

The definitions of FRET probability and stoichiometry used in QuanTI-FRET are mathematically equivalent to what was introduced previously by Chen et al. \cite{chen_measurement_2006} ($\gamma^M \equiv G$ and $\beta^X \equiv 1/(G \cdot k)$) and Lee et al. \cite{lee_accurate_2005} ($\gamma^M \equiv \gamma$ and $\beta^X \equiv \beta$). Therefore, we compared the performances of QuanTI-FRET to these two particular other methods. In the work by Chen et al.\cite{chen_measurement_2006}, the physical origin of the parameters was not described in detail as $\gamma^M$ was already introduced by Zal and Gascoigne \cite{zal_photobleaching-corrected_2004} and the second parameter, $k$, was 
rationally defined from the $\gamma^M$-corrected intensities to account for the stoichiometry. The proposed calibration was achieved in two separated steps.
First, two constructs with defined and well-separated FRET efficiencies were needed to determine $\gamma^M$ (a.k.a $G$). Second, a FRET standard with known stoichiometry was measured to calculate the other parameter, $k$, using $G$ determined in step 1. In Chen's work, the calibration was achieved by imaging the FRET standards C5V and CTV, where the linker $T$ is the 229 amino-acid TRAF domain of the TRAF2 protein \cite{thaler_quantitative_2005}. However, the observation of the 3D representation of all the standards, including CTV, imaged in the present work, shows that CTV does not lie on the same plane as C5V, C17V and C32V (Supplementary Fig.S5). This is also visible in the E-S 2D histogram where the CTV cloud is tilted (Supplementary Fig.S5). These observations are in agreement with the later work of Koushik and Vogel \cite{koushik_energy_2008} and demonstrate the utility of the 3D representation of the fluorescence intensities as well as the E-S 2D histogram to proofread the quality of the experimental data.
The analysis of the experimental dataset [A] with Chen's method gave results close to the QuanTI-FRET method: $G=2.19 \pm 0.02$ to compare with $\gamma^M=2.10 \pm 0.02$ and $1/(G \cdot k) = 1.135 \pm 0.005$ to compare with $\beta^X=1.167 \pm 0.008$ (see Table \ref{table1}). However, the analysis of the second dataset [B] gave different results between the two methods: $G=3.63 \pm 1$ to compare with $\gamma^M=1.35 \pm 0.07$  and $1/(G \cdot k) = 1.02 \pm 7$ to compare with $\beta^X=2.03 \pm 0.07$ yielding less reliable FRET probabilities (respectively 16\%, 24\% and 30\% for C32V, C17V and C5V). This discrepancy results from the dataset being less homogeneous and the limited number of cell-containing pixels where the two-step calibration of $G$ and $k$ is less robust than the single-step fit of the QuanTi-FRET method.
In the work of Lee et al.\cite{lee_accurate_2005}, the calibration consists of first calculating $E_{raw}$ and $S_{raw}$ with only spectral crosstalk corrections and then fitting the linear relation between $1/S_{raw}$ and $E_{raw}$, hereby assuming a 1:1 stoichiometry (see Supplementary Information). This method yielded very similar results to QuanTI-FRET: $\gamma = 2.37 \pm 0.05$ to compare with $\gamma^M=2.10 \pm 0.02$ and $\beta=1.13 \pm 0.01$ to compare with $\beta^X=1.167 \pm 0.008$, resulting in FRET values slightly lower for the FRET standards ($\Delta E = 3\%$). The second dataset ([B]) was also used to test Lee's method leading to a decrease in the FRET values of $\Delta E = 8\%$ with a relative difference of 11\% for $\beta^X$ and 28\% for $\gamma^M$. The correction factors and resulting FRET for the three FRET standards are summarized in Table \ref{table1}. The average FRET probabilities are in very good agreement between QuanTI-FRET and Chen's methods, a systematic difference of about 3\% is observed with Lee's method. The three methods can all be considered as quantitative.

To further test the relative robustness of the three methods, a systematic bootstrap testing on experimental data ([A] with C5V, C17V and C32V) was performed. The whole experimental dataset was randomly divided to produce artificially smaller datasets and give access to statistical errors on the correction factors determination (as given so far). The standard deviation of $\gamma^M$ was around 0.12 (QuanTI-FRET and Chen's) and 0.23 (Lee's) for the minimum tested sample sizes between 1000 and 1300 points. The standard deviation of $\beta^X$ was found to be around 0.04 (QuanTI-FRET and Chen's) and 0.07 (Lee's) for the same range of sample sizes. Over the whole range of sample sizes (from $10^3$ to $10^5$), the standard deviation of both correction factors obtained by Lee's method remained larger than the ones obtained by Chen's and QuanTI-FRET (see Supplementary Fig.S6). This analysis demonstrates that Lee's method is less robust to dataset length, probably due to the fitting of $1/S$ which diverges for small $S$ values. 

A different test was performed by reducing the FRET range of the calibration dataset by taking alternatively only two standards (C5V-C17V, C17V-C32V and C5V-C32V) into account. In this case, Chen's method was not valid anymore for C5V-C17V and C17V-C32V couples resulting in relative variations of 76\% for $G$ and 38\% for $\beta^X$ (see Supplementary Fig.S6). Indeed, Chen's method relies purely on the comparison between the average intensities of two populations, the uncertainty grows as the FRET distance decreases. QuanTI-FRET and Lee's methods, by fitting the total distribution, perform well in this bench test (relative variations of 14\% and 22\% for $\gamma^M$ respectively with QuanTI-FRET and Chen's, and 7\% and 12\% respectively for $\beta^X$, see Supplementary Fig.S5).

All in all, even if the three methods are quantitative in the best case scenario, QuanTI-FRET was demonstrated to be more robust to dataset dispersity, length and FRET range. The single-step calibration in a 3D {$I_{DD}, I^{corr}_{DA}, I_{AA}$} representation, on a continuous distribution of FRET efficiencies allows for the calibration on-the-fly of the sample of interest itself, provided a defined stoichiometry and a distribution of FRET efficiencies in the range of the bench test (at least 5 \%). Taking inspiration from single-molecule literature, we can further exploit stoichiometry to provide a quality check of the experimental data and thereby filter the resulting FRET images.

\section*{Conclusion}
Building upon the previous contributions from live-cell and single-molecule FRET experiments, we present a new framework allowing for quantitative FRET imaging in living cells with a simple multi-channel epifluorescence microscope. Here, we demonstrated the consistency of the method on two different microscopy systems in different laboratories. The QuanTI-FRET method does not require specific instrument for determining spectra or lifetime nor specific hardware development. Image-splitting devices and LED excitation are now commercially available and allow for the same image acquisition protocols as the experimental system used in this work. The QuanTI-FRET calibration does not require acceptor photobleaching, purified proteins or known FRET samples. The only requirement is a known stoichiometry sample (as other quantitative methods) with a broad FRET distribution, which can be obtained directly from the FRET construct of interest (intramolecular-FRET-based biosensors for instance). Nevertheless, an independent calibration using FRET standards is recommended as it allows one to evaluate FRET efficiency and stoichiometry independently. The QuanTI-FRET method was demonstrated to be quantitative and robust, with the additional benefit of having an inherent data quality check.

\section*{Methods}

\subsubsection*{Cells and plasmids}
All plasmids were gifts from Steven Vogel: C5V (Addgene plasmid \# 26394), C17V (Addgene plasmid \# 26395), C32V (Addgene plasmid \# 26396), mVenus N1 (Addgene plasmid \# 27793), mCerulean C1 (Addgene plasmid \# 27796), VCV (Addgene plasmid \# 27788), CVC (Addgene plasmid \# 27809) and CTV (Addgene plasmid \# 27803). Plasmids were amplified in \textit{E.Coli} (DH5$\alpha$) and purified using the NucleoBond\textregistered Xtra kit from Macherey-Nagel GmbH (http://www.mn-net.com). Hela cells were cultured in Dulbecco's Modified Eagle Medium high glucose supplemented with Foetal Bovine Serum (10\%), GlutaMAX\textsuperscript{TM} (Gibco\textsuperscript{TM}) and Penicillin/ Streptomycin (1\%). Cells were transfected with Lipofectamine\textregistered  2000 (Invitrogen\textsuperscript{TM}) and Opti-MEM (Gibco\textsuperscript{TM}), then incubated in Fluorobrite DMEM medium (Gibco\textsuperscript{TM}) overnight and finally imaged in Leibovitz's L-15 medium (Gibco\textsuperscript{TM}) without phenol red.

\subsubsection*{Microscopic image acquisition, Grenoble, setup [A]}
Imaging was done with a system based on an Olympus IX83 body equipped with a home-made image splitting coupled to a sCMOS camera (ORCA Flash V2, Hamamatsu) as sketched in Fig.\ref{Fig1}. Excitation was done by a supercontinuum white laser (Fianium) coupled to a high power AOTF (Fianium), which was controlled through an FPGA-RT unit (National Instruments) coded with Labview. This unit synchronized the alternated laser excitation with the camera acquisition. Images were acquired at $37^{\circ}$C with Micromanager and a 40x objective. The donor fluorophore was excited at 442nm, the acceptor at 515nm. The fluorescence emission was first separated from the excitation via a triple line beamsplitter (Brightline R442/514/561 Semrock) in the microscope body.  The fluorescence emission was further splitted with a beamsplitter at 510nm (Chroma) and filtered with a 475/50 filter (BrightLine HC, Semrock) for the donor channel and a 519/LP longpass filter (BrightLine HC, Semrock) for the acceptor channel. Hence, in two camera snapshots, four images were obtained with all combinations of donor/acceptor excitation and donor/acceptor emission.

\subsubsection*{Microscopic image acquisition, Munich, setup [B]}
Images were acquired on a Nikon Eclipse Ti microscope with home-built excitation and detection pathways. A 100x oil immersion objective (Apo-TIRF 100x Oil/NA 1.49, Nikon) was used for all measurements. Samples were excited with 445nm (MLD, Cobolt) and 514nm (Fandango, Cobolt) diode lasers coupled to an AOTF (PCAOM LFVIS5, Gooch \& Housego) controlled by a FPGA unit (cRIO-9074, National Instruments). The fluorescence emission was separated from excitation pathway with a triple line 445/514/594 beamsplitter. The donor and acceptor emission were separated using an additional 514LP beamsplitter and were then spectrally filtered using 480/40 and 555/55 bandpass filters respectively before being detected on separate EMCCD cameras (DU-897, Andor). Each cell was excited for 300 ms at 445 nm followed by 300 ms at 514 nm. The camera exposure was synchronized to laser excitation through the FPGA unit and a self-written Labview program. This produced four images over two exposure periods capturing donor and acceptor emissions at each excitation wavelength.

\subsubsection*{Image analysis}
All the image analysis calculations were coded in Python, figures and plots were done in Python except for the boxplots obtained with PlotofPlots \cite{postma_plotsofdataweb_2019}. Raw fluorescence images were pre-treated by substracting the dark count of the camera and flattened by dividing with a fluorescence image obtained from a uniform fluorescent sample (Chroma slide). An essential step is then the registration between the two channels obtained on each half of the camera or between cameras. Brightfield images of beads randomly and densely spread on a coverslip were used for calibration. By calculating the image cross-correlations in local regions of the image between the two channels, a displacement map was obtained and hence a transformation matrix was calculated (accounting for translation, rotation, shear and magnification). This transformation matrix was systematically applied to $I_{DD}$ to match $I_{DA}$ and $I_{AA}$ before any calculation. Calibration of the system with QuanTI-FRET was done as explained in the main text. Visualization of the 3D fit was done in Paraview to explore all view angles. All calculations were done pixelwise. Parameters for the weighted gaussian filter are chosen as for gaussian filtering depending on the pixel intensity. Here, the spatial filtering is principally used to filter out pixels with an aberrant stoichiometry, \textit{i.e.} $S$ larger than 0.6 or smaller than 0.4 as estimated from the S-E histograms.  The spatial gaussian enveloppe is designed to avoid adding noise in this operation, as $S$ is subjected to stochastic pixel-to-pixel noise as $E$.

The data that support the findings of this study are available from the corresponding author upon reasonable request.

\bibliography{ArticleFRET}

\begin{thebibliography}{10}
\urlstyle{rm}
\expandafter\ifx\csname url\endcsname\relax
  \def\url#1{\texttt{#1}}\fi
\expandafter\ifx\csname urlprefix\endcsname\relax\def\urlprefix{URL }\fi
\expandafter\ifx\csname doiprefix\endcsname\relax\def\doiprefix{DOI: }\fi
\providecommand{\bibinfo}[2]{#2}
\providecommand{\eprint}[2][]{\url{#2}}

\bibitem{ha_single-molecule_1999}
\bibinfo{author}{Ha, T.} \emph{et~al.}
\newblock \bibinfo{journal}{\bibinfo{title}{Single-molecule fluorescence
  spectroscopy of enzyme conformational dynamics and cleavage mechanism}}.
\newblock {\emph{\JournalTitle{Proc. Natl. Acad. Sci. USA}}} \bibinfo{pages}{6}
  (\bibinfo{year}{1999}).

\bibitem{weiss_fluorescence_1999}
\bibinfo{author}{Weiss, S.}
\newblock \bibinfo{journal}{\bibinfo{title}{Fluorescence {Spectroscopy} of
  {Single} {Biomolecules}}}.
\newblock {\emph{\JournalTitle{Science}}} \textbf{\bibinfo{volume}{283}},
  \bibinfo{pages}{1676--1683}, \doiprefix\url{10.1126/science.283.5408.1676}
  (\bibinfo{year}{1999}).

\bibitem{erickson_preassociation_2001}
\bibinfo{author}{Erickson, M.~G.}, \bibinfo{author}{Alseikhan, B.~a.},
  \bibinfo{author}{Peterson, B.~Z.} \& \bibinfo{author}{Yue, D.~T.}
\newblock \bibinfo{journal}{\bibinfo{title}{Preassociation of calmodulin with
  voltage-gated {Ca} 2+ channels revealed by {FRET} in single living cells}}.
\newblock {\emph{\JournalTitle{Neuron}}} \textbf{\bibinfo{volume}{31}},
  \bibinfo{pages}{973--985}, \doiprefix\url{10.1016/S0896-6273(01)00438-X}
  (\bibinfo{year}{2001}).

\bibitem{zhang_genetically_2001}
\bibinfo{author}{Zhang, J.}, \bibinfo{author}{Ma, Y.}, \bibinfo{author}{Taylor,
  S.~S.} \& \bibinfo{author}{Tsien, R.~Y.}
\newblock \bibinfo{journal}{\bibinfo{title}{Genetically encoded reporters of
  protein kinase {A} activity reveal impact of substrate tethering}}.
\newblock {\emph{\JournalTitle{Proceedings of the National Academy of
  Sciences}}} \textbf{\bibinfo{volume}{98}}, \bibinfo{pages}{14997--15002},
  \doiprefix\url{10.1073/pnas.211566798} (\bibinfo{year}{2001}).

\bibitem{ting_genetically_2001}
\bibinfo{author}{Ting, A.~Y.}, \bibinfo{author}{Kain, K.~H.},
  \bibinfo{author}{Klemke, R.~L.} \& \bibinfo{author}{Tsien, R.~Y.}
\newblock \bibinfo{journal}{\bibinfo{title}{Genetically encoded fluorescent
  reporters of protein tyrosine kinase activities in living cells}}.
\newblock {\emph{\JournalTitle{Proceedings of the National Academy of
  Sciences}}} \textbf{\bibinfo{volume}{98}}, \bibinfo{pages}{15003--15008},
  \doiprefix\url{10.1073/pnas.211564598} (\bibinfo{year}{2001}).

\bibitem{pertz_spatiotemporal_2006}
\bibinfo{author}{Pertz, O.}, \bibinfo{author}{Hodgson, L.},
  \bibinfo{author}{Klemke, R.~L.} \& \bibinfo{author}{Hahn, K.~M.}
\newblock \bibinfo{journal}{\bibinfo{title}{Spatiotemporal dynamics of {RhoA}
  activity in migrating cells.}}
\newblock {\emph{\JournalTitle{Nature}}} \textbf{\bibinfo{volume}{440}},
  \bibinfo{pages}{1069--72}, \doiprefix\url{10.1038/nature04665}
  (\bibinfo{year}{2006}).

\bibitem{miyawaki_fluorescent_1997}
\bibinfo{author}{Miyawaki, A.} \emph{et~al.}
\newblock \bibinfo{journal}{\bibinfo{title}{Fluorescent indicators for {Ca} 2 +
  based on green fluorescent proteins and calmodulin}}.
\newblock {\emph{\JournalTitle{Nature}}} \textbf{\bibinfo{volume}{388}},
  \bibinfo{pages}{882--887} (\bibinfo{year}{1997}).

\bibitem{grashoff_measuring_2010}
\bibinfo{author}{Grashoff, C.} \emph{et~al.}
\newblock \bibinfo{journal}{\bibinfo{title}{Measuring mechanical tension across
  vinculin reveals regulation of focal adhesion dynamics.}}
\newblock {\emph{\JournalTitle{Nature}}} \textbf{\bibinfo{volume}{466}},
  \bibinfo{pages}{263--6}, \doiprefix\url{10.1038/nature09198}
  (\bibinfo{year}{2010}).

\bibitem{meng_fluorescence_2008}
\bibinfo{author}{Meng, F.}, \bibinfo{author}{Suchyna, T.~M.} \&
  \bibinfo{author}{Sachs, F.}
\newblock \bibinfo{journal}{\bibinfo{title}{A fluorescence energy
  transfer-based mechanical stress sensor for specific proteins in situ:
  {Mechanical} stress sensor}}.
\newblock {\emph{\JournalTitle{FEBS Journal}}} \textbf{\bibinfo{volume}{275}},
  \bibinfo{pages}{3072--3087}, \doiprefix\url{10.1111/j.1742-4658.2008.06461.x}
  (\bibinfo{year}{2008}).

\bibitem{ringer_multiplexing_2017}
\bibinfo{author}{Ringer, P.} \emph{et~al.}
\newblock \bibinfo{journal}{\bibinfo{title}{Multiplexing molecular tension
  sensors reveals piconewton force gradient across talin-1}}.
\newblock {\emph{\JournalTitle{Nature Methods}}} \textbf{\bibinfo{volume}{14}},
  \bibinfo{pages}{1090--1096}, \doiprefix\url{10.1038/nmeth.4431}
  (\bibinfo{year}{2017}).

\bibitem{padilla-parra_fret_2012}
\bibinfo{author}{Padilla-Parra, S.} \& \bibinfo{author}{Tramier, M.}
\newblock \bibinfo{journal}{\bibinfo{title}{{FRET} microscopy in the living
  cell: different approaches, strengths and weaknesses.}}
\newblock {\emph{\JournalTitle{Bioessays}}} \textbf{\bibinfo{volume}{34}},
  \bibinfo{pages}{369--76}, \doiprefix\url{10.1002/bies.201100086}
  (\bibinfo{year}{2012}).

\bibitem{chen_characterization_2007}
\bibinfo{author}{Chen, Y.}, \bibinfo{author}{Mauldin, J.~P.},
  \bibinfo{author}{Day, R.~N.} \& \bibinfo{author}{Periasamy, A.}
\newblock \bibinfo{journal}{\bibinfo{title}{Characterization of spectral {FRET}
  imaging microscopy}}.
\newblock {\emph{\JournalTitle{Journal of Microscopy}}}
  \textbf{\bibinfo{volume}{228}}, \bibinfo{pages}{139--152}
  (\bibinfo{year}{2007}).

\bibitem{wlodarczyk_analysis_2008}
\bibinfo{author}{Wlodarczyk, J.} \emph{et~al.}
\newblock \bibinfo{journal}{\bibinfo{title}{Analysis of {FRET} {Signals} in the
  {Presence} of {Free} {Donors} and {Acceptors}}}.
\newblock {\emph{\JournalTitle{Biophysical Journal}}}
  \textbf{\bibinfo{volume}{94}}, \bibinfo{pages}{986--1000},
  \doiprefix\url{10.1529/biophysj.107.111773} (\bibinfo{year}{2008}).

\bibitem{arsenovic_sensorfret:_2017}
\bibinfo{author}{Arsenovic, P.~T.}, \bibinfo{author}{Mayer, C.~R.} \&
  \bibinfo{author}{Conway, D.~E.}
\newblock \bibinfo{journal}{\bibinfo{title}{{SensorFRET}: {A} {Standardless}
  {Approach} to {Measuring} {Pixel}-based {Spectral} {Bleed}-through and {FRET}
  {Efficiency} using {Spectral} {Imaging}}}.
\newblock {\emph{\JournalTitle{Scientific Reports}}}
  \textbf{\bibinfo{volume}{7}}, \bibinfo{pages}{15609},
  \doiprefix\url{10.1038/s41598-017-15411-8} (\bibinfo{year}{2017}).

\bibitem{berney_fret_2003}
\bibinfo{author}{Berney, C.} \& \bibinfo{author}{Danuser, G.}
\newblock \bibinfo{journal}{\bibinfo{title}{{FRET} or no {FRET}: a quantitative
  comparison.}}
\newblock {\emph{\JournalTitle{Biophys. J.}}} \textbf{\bibinfo{volume}{84}},
  \bibinfo{pages}{3992--4010}, \doiprefix\url{10.1016/S0006-3495(03)75126-1}
  (\bibinfo{year}{2003}).

\bibitem{zeug_quantitative_2012}
\bibinfo{author}{Zeug, A.}, \bibinfo{author}{Woehler, A.},
  \bibinfo{author}{Neher, E.} \& \bibinfo{author}{Ponimaskin, E.~G.}
\newblock \bibinfo{journal}{\bibinfo{title}{Quantitative intensity-based {FRET}
  approaches - {A} comparative snapshot}}.
\newblock {\emph{\JournalTitle{Biophys. J.}}} \textbf{\bibinfo{volume}{103}},
  \bibinfo{pages}{1821--1827}, \doiprefix\url{10.1016/j.bpj.2012.09.031}
  (\bibinfo{year}{2012}).

\bibitem{dahan_ratiometric_1999}
\bibinfo{author}{Dahan, M.} \emph{et~al.}
\newblock \bibinfo{journal}{\bibinfo{title}{Ratiometric measurement and
  identification of single diffusing molecules}}.
\newblock {\emph{\JournalTitle{Chemical Physics}}}
  \textbf{\bibinfo{volume}{247}}, \bibinfo{pages}{85--106},
  \doiprefix\url{10.1016/S0301-0104(99)00132-9} (\bibinfo{year}{1999}).

\bibitem{kapanidis_fluorescence-aided_2004}
\bibinfo{author}{Kapanidis, A.~N.} \emph{et~al.}
\newblock \bibinfo{journal}{\bibinfo{title}{Fluorescence-aided molecule
  sorting: analysis of structure and interactions by alternating-laser
  excitation of single molecules.}}
\newblock {\emph{\JournalTitle{Proc. Natl. Acad. Sci. U. S. A.}}}
  \textbf{\bibinfo{volume}{101}}, \bibinfo{pages}{8936--8941},
  \doiprefix\url{10.1073/pnas.0401690101} (\bibinfo{year}{2004}).

\bibitem{youvan_fluorescence_1997}
\bibinfo{author}{Youvan, D.~C.} \emph{et~al.}
\newblock \bibinfo{journal}{\bibinfo{title}{Fluorescence {Imaging}
  {Micro}-{Spectrophotometer} ({FIMS})}}.
\newblock {\emph{\JournalTitle{Biotechnology et alia}}}
  \textbf{\bibinfo{volume}{1}}, \bibinfo{pages}{1--16} (\bibinfo{year}{1997}).

\bibitem{gordon_quantitative_1998}
\bibinfo{author}{Gordon, G.~W.}, \bibinfo{author}{Berry, G.},
  \bibinfo{author}{Liang, X.~H.}, \bibinfo{author}{Levine, B.} \&
  \bibinfo{author}{Herman, B.}
\newblock \bibinfo{journal}{\bibinfo{title}{Quantitative fluorescence resonance
  energy transfer measurements using fluorescence microscopy.}}
\newblock {\emph{\JournalTitle{Biophys. J.}}} \textbf{\bibinfo{volume}{74}},
  \bibinfo{pages}{2702--2713}, \doiprefix\url{10.1016/S0006-3495(98)77976-7}
  (\bibinfo{year}{1998}).

\bibitem{zal_photobleaching-corrected_2004}
\bibinfo{author}{Zal, T.} \& \bibinfo{author}{Gascoigne, N. R.~J.}
\newblock \bibinfo{journal}{\bibinfo{title}{Photobleaching-corrected {FRET}
  efficiency imaging of live cells.}}
\newblock {\emph{\JournalTitle{Biophys. J.}}} \textbf{\bibinfo{volume}{86}},
  \bibinfo{pages}{3923--39}, \doiprefix\url{10.1529/biophysj.103.022087}
  (\bibinfo{year}{2004}).

\bibitem{hochreiter_advanced_2019}
\bibinfo{author}{Hochreiter, B.}, \bibinfo{author}{Kunze, M.},
  \bibinfo{author}{Moser, B.} \& \bibinfo{author}{Schmid, J.~A.}
\newblock \bibinfo{journal}{\bibinfo{title}{Advanced {FRET} normalization
  allows quantitative analysis of protein interactions including
  stoichiometries and relative affinities in living cells}}.
\newblock {\emph{\JournalTitle{Scientific Reports}}}
  \textbf{\bibinfo{volume}{9}}, \bibinfo{pages}{8233},
  \doiprefix\url{10.1038/s41598-019-44650-0} (\bibinfo{year}{2019}).

\bibitem{hoppe_fluorescence_2002}
\bibinfo{author}{Hoppe, A.}, \bibinfo{author}{Christensen, K.} \&
  \bibinfo{author}{Swanson, J.~a.}
\newblock \bibinfo{journal}{\bibinfo{title}{Fluorescence resonance energy
  transfer-based stoichiometry in living cells.}}
\newblock {\emph{\JournalTitle{Biophys. J.}}} \textbf{\bibinfo{volume}{83}},
  \bibinfo{pages}{3652--64}, \doiprefix\url{10.1016/S0006-3495(02)75365-4}
  (\bibinfo{year}{2002}).

\bibitem{thaler_quantitative_2005}
\bibinfo{author}{Thaler, C.}, \bibinfo{author}{Koushik, S.~V.},
  \bibinfo{author}{Blank, P.~S.} \& \bibinfo{author}{Vogel, S.~S.}
\newblock \bibinfo{journal}{\bibinfo{title}{Quantitative {Multiphoton}
  {Spectral} {Imaging} and {Its} {Use} for {Measuring} {Resonance} {Energy}
  {Transfer}}}.
\newblock {\emph{\JournalTitle{Biophys. J.}}} \textbf{\bibinfo{volume}{89}},
  \bibinfo{pages}{2736--2749}, \doiprefix\url{10.1529/biophysj.105.061853}
  (\bibinfo{year}{2005}).

\bibitem{lee_accurate_2005}
\bibinfo{author}{Lee, N.~K.} \emph{et~al.}
\newblock \bibinfo{journal}{\bibinfo{title}{Accurate {FRET} measurements within
  single diffusing biomolecules using alternating-laser excitation.}}
\newblock {\emph{\JournalTitle{Biophys. J.}}} \textbf{\bibinfo{volume}{88}},
  \bibinfo{pages}{2939--53}, \doiprefix\url{10.1529/biophysj.104.054114}
  (\bibinfo{year}{2005}).

\bibitem{koushik_cerulean_2006}
\bibinfo{author}{Koushik, S.~V.}, \bibinfo{author}{Chen, H.},
  \bibinfo{author}{Thaler, C.}, \bibinfo{author}{Iii, H. L.~P.} \&
  \bibinfo{author}{Vogel, S.~S.}
\newblock \bibinfo{journal}{\bibinfo{title}{Cerulean , {Venus} , and
  {VenusY}67c {FRET} {Reference} {Standards}}}.
\newblock {\emph{\JournalTitle{Biophys. J.}}} \textbf{\bibinfo{volume}{91}},
  \bibinfo{pages}{L99--L101}, \doiprefix\url{10.1529/biophysj.106.096206}
  (\bibinfo{year}{2006}).

\bibitem{hellenkamp_precision_2018}
\bibinfo{author}{Hellenkamp, B.} \emph{et~al.}
\newblock \bibinfo{journal}{\bibinfo{title}{Precision and accuracy of
  single-molecule {FRET} measurements—a multi-laboratory benchmark study}}.
\newblock {\emph{\JournalTitle{Nature Methods}}} \textbf{\bibinfo{volume}{15}},
  \bibinfo{pages}{669--676}, \doiprefix\url{10.1038/s41592-018-0085-0}
  (\bibinfo{year}{2018}).

\bibitem{ha_probing_1996}
\bibinfo{author}{Ha, T.} \emph{et~al.}
\newblock \bibinfo{journal}{\bibinfo{title}{Probing the interaction between two
  single molecules: fluorescence resonance energy transfer between a single
  donor and a single acceptor.}}
\newblock {\emph{\JournalTitle{Proceedings of the National Academy of
  Sciences}}} \textbf{\bibinfo{volume}{93}}, \bibinfo{pages}{6264--6268},
  \doiprefix\url{10.1073/pnas.93.13.6264} (\bibinfo{year}{1996}).

\bibitem{chen_measurement_2006}
\bibinfo{author}{Chen, H.}, \bibinfo{author}{Puhl, H.~L.},
  \bibinfo{author}{Koushik, S.~V.}, \bibinfo{author}{Vogel, S.~S.} \&
  \bibinfo{author}{Ikeda, S.~R.}
\newblock \bibinfo{journal}{\bibinfo{title}{Measurement of {FRET} {Efficiency}
  and {Ratio} of {Donor} to {Acceptor} {Concentration} in {Living} {Cells}}}.
\newblock {\emph{\JournalTitle{Biophysical Journal}}}
  \textbf{\bibinfo{volume}{91}}, \bibinfo{pages}{L39--L41},
  \doiprefix\url{10.1529/biophysj.106.088773} (\bibinfo{year}{2006}).

\bibitem{frommer_genetically_2009}
\bibinfo{author}{Frommer, W.~B.}, \bibinfo{author}{Davidson, M.~W.} \&
  \bibinfo{author}{Campbell, R.~E.}
\newblock \bibinfo{journal}{\bibinfo{title}{Genetically encoded biosensors
  based on engineered fluorescent proteins}}.
\newblock {\emph{\JournalTitle{Chemical Society Reviews}}}
  \textbf{\bibinfo{volume}{38}}, \bibinfo{pages}{2833},
  \doiprefix\url{10.1039/b907749a} (\bibinfo{year}{2009}).

\bibitem{sizaire_fret-based_2017}
\bibinfo{author}{Sizaire, F.} \& \bibinfo{author}{Tramier, M.}
\newblock \bibinfo{title}{{FRET}-based biosensors: genetically encoded tools to
  track kinase activity in living cells}.
\newblock In \emph{\bibinfo{booktitle}{Protein phosphorylation}},
  \bibinfo{pages}{179} (\bibinfo{publisher}{IntechOpen}, \bibinfo{year}{2017}).

\bibitem{koushik_energy_2008}
\bibinfo{author}{Koushik, S.~V.} \& \bibinfo{author}{Vogel, S.~S.}
\newblock \bibinfo{journal}{\bibinfo{title}{Energy migration alters the
  fluorescence lifetime of {Cerulean}: implications for fluorescence lifetime
  imaging {Forster} resonance energy transfer measurements}}.
\newblock {\emph{\JournalTitle{Journal of Biomedical Optics}}}
  \textbf{\bibinfo{volume}{13}}, \bibinfo{pages}{031204},
  \doiprefix\url{10.1117/1.2940367} (\bibinfo{year}{2008}).

\bibitem{postma_plotsofdataweb_2019}
\bibinfo{author}{Postma, M.} \& \bibinfo{author}{Goedhart, J.}
\newblock \bibinfo{journal}{\bibinfo{title}{{PlotsOfData}—{A} web app for
  visualizing data together with their summaries}}.
\newblock {\emph{\JournalTitle{PLOS Biology}}} \textbf{\bibinfo{volume}{17}},
  \bibinfo{pages}{e3000202}, \doiprefix\url{10.1371/journal.pbio.3000202}
  (\bibinfo{year}{2019}).

\end{thebibliography}

\section*{Acknowledgements}
The authors are grateful to S. Pinel and N. Scaramozzino (M2Bio platform) for advices and help in plasmid amplification and purification. We thank B. Arnal and I. Wang for fruitful discussions and numerical advices. P.Moreau is acknowledged for his technical participation in building and maintaining the setup. This work was funded by the Agence Nationale de la Recherche (ANR, grant $n^{\circ}$ ANR-13-PDOC-0022-01), and supported by the Université Grenoble Alpes (UGA, AGIR-POLE program 2015, project ACTSUB). C.M.B., A.C. and A.D. are part of the GDR 3070 CellTiss. D.C.L. gratefully acknowledges the financial support of the Deutsche Forschungsgemeinschaft (DFG) via the collaborative research center (SFB1035, Project A11) and the Ludwig-Maximilians-Universität through the Center for NanoScience (CeNS) and the BioImaging Network (BIN). CAR and CO are supported by FRM and ANR (CE17 CE13 0022 01).

\section*{Author contributions statement}
A.D. conceived the experiments and the theory. D.C.L contributed to the concept. C.O and C.A-R contributed to the design of the biological protocols. A.C., C.M.B, C.Q. conducted the experiments. A.D and A.C analyzed the results. F.W. and C.M.B built and interfaced the hardware. A.D drafted the manuscript. All authors reviewed the manuscript and approved the final version.

\section*{Additional information}
The author(s) declare no competing interests.

\end{document}